\documentclass[12pt]{iopart}
\usepackage{iopams} 
\usepackage{cite}
\usepackage{graphicx}
\newcommand{\pster}{\pi^\ast}
\newcommand{\wigner}[6]{\left(\begin{array}{ccc} #1 & #2 & #3 \\ #4 & #5 & #6 \end{array}\right)} 
\begin{document}
\title{Quadrupole collective variables in the natural Cartan-Weyl basis}
\author{S.~De~Baerdemacker, K.~Heyde and V.~Hellemans}
\address{Universiteit Gent, Vakgroep Subatomaire en Stralingsfysica, Proeftuinstraat 86, B-9000 Gent, Belgium}
\ead{stijn.debaerdemacker@ugent.be}
\begin{abstract}
The matrix elements of the quadrupole collective variables, emerging from collective nuclear models, are calculated in the natural Cartan-Weyl basis of $O(5)$ which is a subgroup of a covering $SU(1,1)\times O(5)$ structure.  Making use of an intermediate set method, explicit expressions of the matrix elements are obtained in a pure algebraic way, fixing the $\gamma$-rotational structure of collective quadrupole models.
\end{abstract}
\submitto{\JPA}
\maketitle

\section{Introduction}
%
%
Collective modes of motion play a significant role in the low-energy structure of atomic nuclei.  To account for large quadrupole moments, the spherical shell-model picture needed to be extended to spheroidal deformations \cite{rainwater:50}, due to collective polarization effects induced by single particles moving in the nuclear medium.   As a consequence the nucleus can no longer be regarded as a rigid body, rather a soft object with a surface that can undergo oscillations and rotations in the laboratory framework.  The quantized treatment of these excitations in the intrinsic framework \cite{bohr:52,bohr:53} led to the development of the Bohr Hamiltonian in terms of the intrinsic collective variables $\beta$ and $\gamma$ \cite{eisenberg:87,bohr:98}, corresponding respectively to the degree of axial and triaxial deformation.   The dynamics is determined by the potential contained in the Bohr Hamiltonian, which can either be constructed from a microscopic theory or through phenomenological considerations. When the latter strategy is followed, one can either choose an analytically solvable potential, a topic which has recently gained a considerable amount of interest because of its application to critical points in phase shape transitions \cite{iachello:00,iachello:01,iachello:03}, or a more general expression \cite{eisenberg:87} in terms of the collective variables, determining the surface (\ref{ellipsoid:equation}).  For an overview on (approximative) analytically solvable potentials, we would like to refer the reader to \cite{fortunato:05}.\\
Analytically solvable potentials are intended as benchmarks in order to study more general and complex potentials.  To handle these potentials, one needs to perform a diagonalization, for which a suitable basis is needed.  Within the literature, several methods have been proposed and profoundly discussed.  Pioneering work has been carried out by  B\`es \cite{bes:58}, who determined the explicit $\gamma$-soft wavefunctions through a coupled differential equation method.  Unfortunately, this technique becomes tremendously complicated for spin states, higher than $L=6$.  Therefore, other techniques have been developed, fully exploiting the $SU(1,1)\times O(5)$ structure of the five-dimensional harmonic oscillator.  Nevertheless, complications arise.  Since the Hamiltonian is an angular momentum scalar, the eigenstates automatically possess good quantum numbers $L$ and $M$ of the physical $O(3)\supset O(2)$ chain which does not evolve naturally from the Cartan-Weyl group reduction.  As a consequence one is forced to construct explicit wavefunctions, starting either from basic building blocks \cite{corrigan:76,chacon:76,chacon:77} or from a projective coherent state formalism \cite{gheorghe:78}, constituting an orthonormal basis \cite{szpikowski:80,gozdz:80}.  Nevertheless, the alternative Cartan-Weyl \cite{cartan:1894,wybourne:74}  reduction path may also be followed as it leads to a reduction scheme which is more natural in a mathematical sense, though the physical meaning of the quantum numbers is partially lost.  This strategy was followed by Hecht \cite{hecht:64} to construct fractional parentage coefficients for spin-2 phonons, that were used by the Frankfurt group \cite{eisenberg:87,gneuss:71} in the development of the General Collective model.\\
It is noteworthy that new techniques have been proposed within the last decennium.  First, the vector coherent state formalism \cite{rowe:94a,rowe:94b,rowe:95,turner:06} and much more recently the algebraic tractable model \cite{rowe:04,rowe:05a} were developed, enabling the construction of the quadrupole harmonic oscillator representations, exhibiting good  $O(3)$ angular momentum quantum numbers.\\
In the present paper, the path of the natural Cartan-Weyl reduction is taken.  It will be shown that the matrix elements of the quadrupole variable can be extracted within this basis, without making use of the explicit representations in terms of the collective variables.   In the end, it will turn out that the basic commutation relations of the collective variables suffice to fix the complete structure of the algebra, and furthermore the dynamics of the Hamiltonian.
\section{The collective model}
%
%
Within the framework of the geometrical model, the nucleus is regarded as a liquid drop with a surface $R(\theta,\phi)$ described by a multipole expansion using spherical harmonics in the laboratory system
\begin{equation}\label{ellipsoid:equation}
R(\theta,\phi)=R_0\Big(1+\sum_{\lambda,\mu}\alpha^\ast_{\lambda,\mu}Y_{\lambda,\mu}(\theta,\phi)\Big),
\end{equation}
which defines the set of collective coordinates $\alpha_{\lambda,\mu}$ of multipolarity $\lambda$ and projection $\mu$.  Up to quadrupole deformation, the surface (\ref{ellipsoid:equation}) is restricted to spheroidal deformations determined by the variables $\alpha_{2,\mu}$ which will be abbreviated to $\alpha_\mu$ from here on.  Although the intrinsic surface is unambiguously described by this set of variables, it is convenient to rotate from the laboratory to the intrinsic framework by means of the Euler angles ($\theta_i,i=1,2,3$).  Doing so, the collective variables $(\beta,\gamma)$ are introduced as intrinsic parameters of the ellipsoid, rendering a straightforward interpretation of axial and triaxial deformation \cite{eisenberg:87}.\\
\begin{equation}\label{variables:intrinsicframe}
\alpha_\mu=\beta\left[\cos\gamma D_{\mu 0}^{2\ast}(\theta_i)+\frac{\sin\gamma}{\sqrt{2}}(D_{\mu2}^{2\ast}(\theta_i)+D_{\mu-2}^{2\ast}(\theta_i))\right].
\end{equation}
This set of collective variables is sufficient for the determination of the static properties of nuclear shapes.  To build in the essential quantum mechanical dynamics, canonic conjugate momenta $\pi_{\mu^\prime}$ need to be incorporated.  These must fulfill the standard commutation relations \cite{eisenberg:87}
\begin{equation}\label{variables:commutationrelations}
[\pi_{\mu^\prime},\alpha_\mu]=-i\hbar\delta_{\mu\mu^\prime},\qquad [\pi_{\mu^\prime},\pi_\mu]=0,\qquad [\alpha_{\mu^\prime},\alpha_\mu]=0.
\end{equation}
Note that the variables have become operators though we silently omit the operator symbol to avoid notational overload.\\
To establish the $SU(1,1)\times O(5)$ group structure, it is convenient to introduce the following recoupling formula
\begin{equation}\label{generators:introduction}
\eqalign{(\alpha\cdot\alpha)(\pi^\ast\cdot\pi^\ast)=&(\alpha\cdot\pi^\ast)(\alpha\cdot\pi^\ast)+3i\hbar(\alpha\cdot\pi^\ast)\\&-2([\alpha\pi^\ast]^{(1)}\cdot[\alpha\pi^\ast]^{(1)}+[\alpha\pi^\ast]^{(3)}\cdot[\alpha\pi^\ast]^{(3)}),}
\end{equation}
where the complex conjugate $\pi^\ast$ is introduced to ensure for good angular momentum transformation properties and $T_l\cdot S_l=(-)^l\sqrt{2l+1}[T_l S_l]^{(0)}_0$.\\
The 3 operators $(\alpha\cdot\alpha)$, $(\pi^\ast\cdot\pi^\ast)$ and $(\alpha\cdot\pi^\ast)$ generate the algebra of an $SU(1,1)$ group, which forms a direct product together with the $O(5)$ group, built from the 10 operators $[\alpha\pi^\ast]^{(1)}_M$ and $[\alpha\pi^\ast]^{(3)}_M$.  Whereas the $SU(1,1)$ group is strongly linked to the excitations in the radial variable $\beta$, the $O(5)$ group encompasses the $\gamma$ vibrations coupled to the rotational structure.  In this work, we concentrate on the application of the Cartan-Weyl scheme on the $O(5)$ group, leaving a freedom of choice of a suitable $SU(1,1)$ basis.
\section{The Cartan-Weyl reduction of $O(5)$}
\begin{figure}[!htb]
\includegraphics{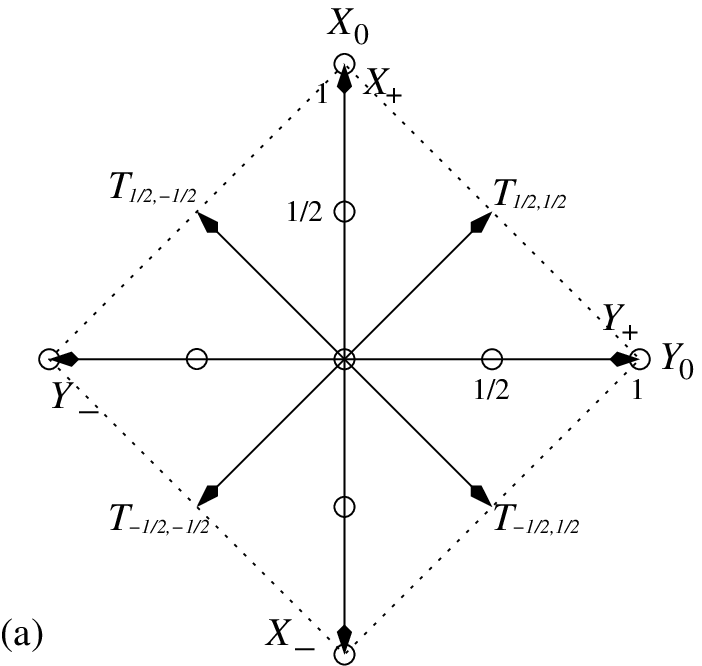}
\includegraphics{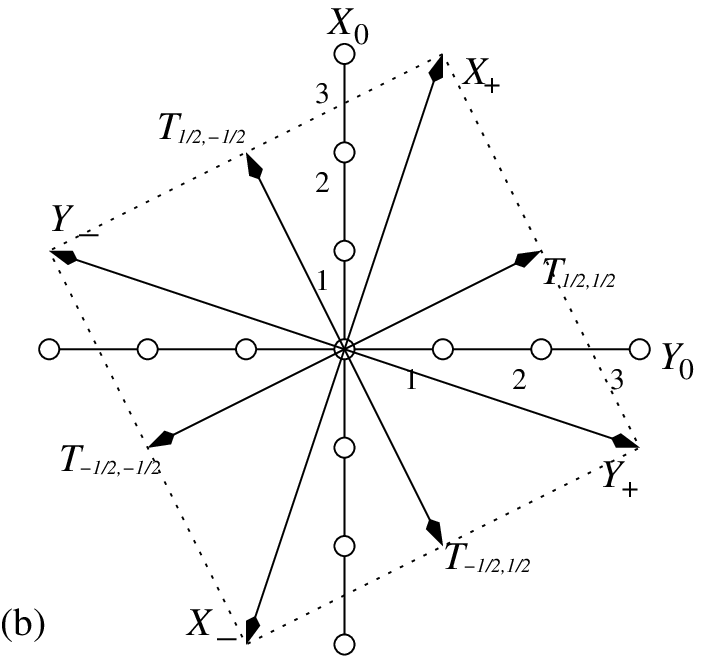}
\caption{The root diagrams of the $O(5)$ algebra in the Cartan-Weyl basis for either the (a) natural $\{X_0,Y_0\}$ or (b) physical $\{L_0,O_0\}$ Cartan subalgebra }\label{figure:doublecartan}
\end{figure}

%
%
The commutation relations of the operators $L_M$ and $O_M$, defined by
\begin{equation}\label{LMOM:definition}
\eqalign{
[\alpha\pi^\ast]^{(1)}_M=\case{i\hbar}{\sqrt{10}}L_M,\\
{}[\alpha\pi^\ast]^{(3)}_M=\case{i\hbar}{\sqrt{10}}O_M,}
\end{equation}
span the algebra of the $O(5)$ group.
\begin{eqnarray}
&[L_m,L_{m^\prime}]=-\sqrt{2}\langle 1m1m^\prime|1m+m^\prime\rangle L_{m+m^\prime},\\{}
&[L_m,O_{m^\prime}]=-2\sqrt{3}\langle 1m3m^\prime|3m+m^\prime\rangle O_{m+m^\prime},\\{}
&[O_m,O_{m^\prime}]=-2\sqrt{7}\langle 3m3m^\prime|1m+m^\prime\rangle L_{m+m^\prime}\nonumber\\{}
&\qquad\qquad\qquad+\sqrt{6}\langle 3m3m^\prime|3m+m^\prime\rangle O_{m+m^\prime}.
\end{eqnarray}
The $M=0$ projections $\{L_0,O_0\}$ form an intuitive choice of the Cartan subalgebra within the set $\{L_m,O_{m^\prime}\}$.  This set has the advantage of incorporating the angular momentum projection operator $L_0$ in the physical group reduction chain $O(5)\supset O(3)\supset O(2)$.  Nevertheless, it is not explicitly contained in the natural Cartan-Weyl reduction chain $O(5)\supset O(4)\cong SU(2)\times SU(2)$ which can be realized through the following rotation \cite{corrigan:76,rowe:94a}
\begin{equation}\label{generators:rotationtocartan}
\eqalign{
X_+=-\case{1}{5}(\sqrt{2}L_{+1}+\sqrt{3}O_{+1}),\qquad & Y_+=-\case{1}{\sqrt{5}}O_{+3},\\
X_-=\case{1}{5}(\sqrt{2}L_{-1}+\sqrt{3}O_{-1}), & Y_-=\case{1}{\sqrt{5}}O_{-3},\\
X_0=\case{1}{10}(L_0+3O_0), & Y_0=\case{1}{10}(3L_0-O_0),\\
T_{\frac{1}{2}\frac{1}{2}}=\case{1}{\sqrt{10}}O_{+2}, & T_{-\frac{1}{2}\frac{1}{2}}=-\case{1}{\sqrt{50}}(\sqrt{3}L_{+1}-\sqrt{2}O_{+1}),\\
T_{-\frac{1}{2}-\frac{1}{2}}=-\case{1}{\sqrt{10}}O_{-2}, &  T_{\frac{1}{2}-\frac{1}{2}}=\case{1}{\sqrt{50}}(\sqrt{3}L_{-1}-\sqrt{2}O_{-1}).}
\end{equation}
The group reduction is immediately clear, as the sets $\{X_0,X_\pm\}$ and $\{Y_0,Y_\pm\}$ both span standard $SU(2)$ algebras.  Furthermore all generators of the one $SU(2)$ algebra commute with all generators of the other.  The commutation relations are given by
\begin{equation}
\eqalign{[X_0,X_\pm]=\pm X_\pm, & [X_+,X_-]=2X_0,\\{}
[Y_0,Y_\pm]=\pm Y_\pm, & [Y_+,Y_-]=2Y_0,\\{}
[X_0,Y_0]=0, & [X_\pm,Y_\pm]=[X_\pm,Y_\mp]=0.}
\end{equation}
So the reduction is $O(5)\supset O(4)\cong SU(2)_X\times SU(2)_Y$.  The non-$O(4)$ operators $T_{\mu\nu}$ can be identified as the 4 components of a bitensor of character $\{\frac{1}{2},\frac{1}{2}\}$ within the $SU(2)\times SU(2)$ scheme, according to Racah's definition \cite{racah:42}.  The index $\mu$ denotes the bitensor component relative to the $SU(2)_X$ group, while $\nu$ is the component with respect to $SU(2)_Y$ 
\begin{equation}
\eqalign{[X_{0},T_{\mu\nu}]=\mu T_{\mu\nu},\\{}
[X_{\pm},T_{\mu\nu}]=\sqrt{(\case{1}{2}\mp\mu)(\case{1}{2}\pm\mu+1)}T_{\mu\pm1\nu},\\{}
[Y_{0},T_{\mu\nu}]=\nu T_{\mu\nu},\\{}
[Y_{\pm},T_{\mu\nu}]=\sqrt{(\case{1}{2}\mp\nu)(\case{1}{2}\pm\nu+1)}T_{\mu\nu\pm1}.}
\end{equation}
The internal commutation relations of the $T$ bitensor completes the Cartan-Weyl structure, 
\begin{table}[!htb]
\begin{tabular}{c|*{4}{c}}
  $*$ & $T_{-\frac{1}{2}-\frac{1}{2}}$ & $T_{\frac{1}{2}-\frac{1}{2}}$ & $T_{-\frac{1}{2}\frac{1}{2}}$ & $T_{\frac{1}{2}\frac{1}{2}}$ \\
 \hline
 $T_{-\frac{1}{2}-\frac{1}{2}}$ & 0 & $\frac{1}{2}Y_{-}$ & $\frac{1}{2}X_{-}$ & $\frac{1}{2}(X_0+Y_0)$\\
 $T_{\frac{1}{2}-\frac{1}{2}}$ & $-\frac{1}{2}Y_{-}$ & 0 & $\frac{1}{2}(X_0-Y_0)$ & $-\frac{1}{2}X_{+}$\\
 $T_{-\frac{1}{2}\frac{1}{2}}$ & $-\frac{1}{2}X_{-}$ & $-\frac{1}{2}(X_0-Y_0)$ & 0 & $-\frac{1}{2}Y_{+}$\\
 $T_{\frac{1}{2}\frac{1}{2}}$ & $-\frac{1}{2}(X_0+Y_0)$ & $\frac{1}{2}X_{+}$ & $\frac{1}{2}Y_{+}$ & 0\\
\end{tabular}
\caption{Multiplication table for the internal commutation relations of the $T$ bitensor.  The multiplication $*$ symbolizes the standard commutation.}\label{table:bitensorcommutation}
\end{table}
 which can be found in table (\ref{table:bitensorcommutation})\\
Once the commutation relations have been determined within the Cartan-Weyl basis, it is instructive to construct the root diagram.  Figure \ref{figure:doublecartan} shows 2 different realizations of the same root diagram, depending on the choice of the Cartan subalgebra.  On the left side (Fig.~\ref{figure:doublecartan}a) a standard root diagram with respect to the $\{X_0,Y_0\}$ Cartan subalgebra is depicted, while on the right side (Fig.~\ref{figure:doublecartan}b), a more physical subalgebra $\{L_0,O_0\}$ is chosen as a reference frame.  The latter framework has a visual advantage, since the projection of the generators on the $L_0$-axis is readily established.  This enhances the insight in the problem of constructing wavefunctions with good angular momentum from the weight diagrams in the Cartan-Weyl basis (see section (\ref{section:rotation})).
\section{Representations of $O(5)$}\label{section:representations}
\begin{figure}[!htb]
\begin{center}
\includegraphics[width=0.5\textwidth]{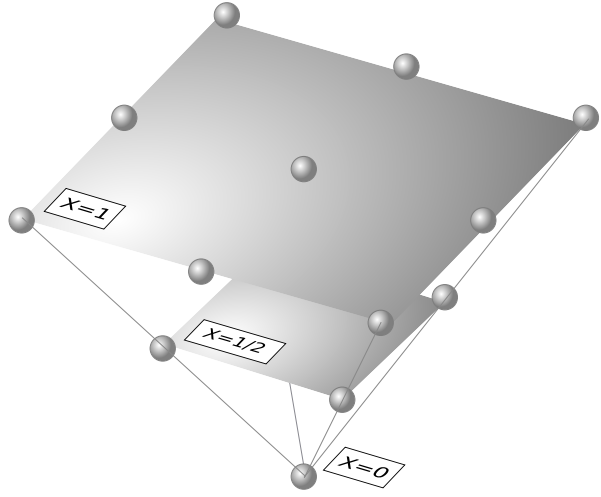}
\end{center}
\caption{Visual interpretation of an $O(5)$ representation with $v=2$.  Every sphere denotes a single basis state. The representation is organized in planes with distinct $X$ quantum number, which contain $(2X+1)^2$ ($M_X,M_Y$) projection states.}\label{figure:3Dmschema}
\end{figure}

%
%
Every subgroup in the group reduction chain provides an associated Casimir operator.  The quadratic Casimir operator of $O(5)$ can be constructed from the Killing form \cite{wybourne:74}
\begin{eqnarray}
\mathcal{C}_2[O(5)]&=\case{1}{5}(L\cdot L+O\cdot O),\\
&=2(X^2+Y^2-2[TT]^{(00)}).\label{casimir:O5}
\end{eqnarray}
with $[TT]^{(00)}$ denoting the scalar Clebsch Gordan coupling with respect to both $SU(2)_X$ and $SU(2)_Y$, and $X^2$ and $Y^2$ the quadratic Casimir operator of the respective $SU(2)$ groups
\begin{eqnarray}
X^2&=X_0^2+\case{1}{2}(X_{+}X_{-}+X_{-}X_{+})\label{casimir:SU2X},\\
Y^2&=Y_0^2+\case{1}{2}(Y_{+}Y_{-}+Y_{-}Y_{+})\label{casimir:SU2Y}.
\end{eqnarray}
Starting from the explicit expressions of the generators (see \ref{appendix:A}) in terms of the collective variables and the canonic conjugate momenta, the following operator identity can be proven
\begin{equation}
X^2-Y^2\equiv0,
\end{equation}
which is true in general for symmetric representations \cite{corrigan:76}.  The consequence of this identity is that we are left with 4 operators that commute among each others, i.e. the quadratic Casimir operator of $O(5)$, the quadratic Casimir operator of $SU(2)_X$ and $SU(2)_Y$ ($X^2\equiv Y^2$) and the Cartan subalgebra $\{X_0,Y_0\}$ which are the respective linear Casimir operators of the $O(2)_X$ and $O(2)_Y$ subgroups.  As a result, we obtain a representation which is determined by 4 independent quantum numbers
\begin{equation}
|vXM_XM_Y\rangle
\end{equation}
with
\begin{eqnarray}
\mathcal{C}_2[O(5)]|vXM_XM_Y\rangle=v(v+3)|vXM_XM_Y\rangle,\\
X^2|vXM_XM_Y\rangle=Y^2|vXM_XM_Y\rangle=X(X+1)|vXM_XM_Y\rangle,\\
X_0|vXM_XM_Y\rangle=M_X|vXM_XM_Y\rangle,\\
Y_0|vXM_XM_Y\rangle=M_Y|vXM_XM_Y\rangle.
\end{eqnarray}
Now that the basis to work in is fixed, we can study the action of the generators as they hop through the representations with fixed quantum number $v$.  Acting with the $O(4)\cong SU(2)_X\times SU(2)_Y$ generators on $|vXM_XM_Y\rangle$ is trivial because of the well-known angular momentum theory
\begin{eqnarray}
X_{\pm}|vXM_XM_Y\rangle&=\sqrt{(X\mp M_X)(X\pm M_X+1)}|vXM_X\pm1,M_Y\rangle,\\
X_{0}|vXM_XM_Y\rangle&=M_X|vXM_XM_Y\rangle,\\
Y_{\pm}|vXM_XM_Y\rangle&=\sqrt{(X\mp M_Y)(X\pm M_Y+1)}|vXM_X,M_Y\pm1\rangle,\\
Y_{0}|vXM_XM_Y\rangle&=M_Y|vXM_XM_Y\rangle.
\end{eqnarray}
The action of $T_{\mu\nu}$ on $|vX(M_X,M_Y)\rangle$ is less trivial, though the bitensorial character of $T$ can be well exploited.  Since $T$ is a $\{\frac{1}{2}\frac{1}{2}\}$ bitensor, it can only connect representations that differ $\frac{1}{2}$ in quantum number $X$
\begin{equation}
\eqalign{
T_{\mu\nu}|vXM_XM_Y\rangle&=a|v,X+\case{1}{2},M_X+\mu,M_Y+\nu\rangle\\
&\qquad+b|v,X-\case{1}{2},M_X+\mu,M_Y+\nu\rangle}.
\end{equation}
The coefficients $a$ and $b$ are not only dependent on $v$ and $X$, but also on the projection quantum numbers $\mu$, $\nu$, $M_X$ and  $M_Y$.  However, these projections can be filtered out by means of the Wigner-Eckart theorem.  As the $SU(2)_X$ forms a direct product with $SU(2)_Y$, we can apply the theorem for both groups, independently from each other.  As a result, the dependency on the projection quantum numbers is completely factored out in the Wigner-$3j$ symbols.  This leaves a double reduced matrix element \footnote{We formally use the single reduced matrix notation in order to express the double reduced matrix, as any confusion between normal and double reduced matrix element is excluded within this work.} to be calculated. 
\begin{eqnarray}
\eqalign{\fl a&=\langle v,X+\case{1}{2},M_X+\mu,M_Y+\nu|T_{\mu\nu}|vXM_XM_Y\rangle \\
\fl &=(-)^{k}\wigner{X+\frac{1}{2}}{\frac{1}{2}}{X}{-M_X-\mu}{\mu}{M_X}\wigner{X+\frac{1}{2}}{\frac{1}{2}}{X}{-M_Y-\nu}{\nu}{M_Y}\langle vX+\frac{1}{2}||T||vX\rangle,}\label{Tgenerators:awignereckart}\\
\eqalign{\fl b&=\langle v,X-\case{1}{2},M_X+\mu,M_Y+\nu|T_{\mu\nu}|vXM_XM_Y\rangle \\
\fl &=(-)^{k}\wigner{X-\frac{1}{2}}{\frac{1}{2}}{X}{-M_X-\mu}{\mu}{M_X}\wigner{X-\frac{1}{2}}{\frac{1}{2}}{X}{-M_Y-\nu}{\nu}{M_Y}\langle vX-\frac{1}{2}||T||vX\rangle,}\label{Tgenerators:bwignereckart}
\end{eqnarray}
with $k=2X+1-M_X-M_Y-\mu-\nu$.\\
In order to calculate the double reduced matrix element, we have 2 types of expressions at hand.  On the one hand the internal commutation relations of the $T$ bitensor (see Table~\ref{table:bitensorcommutation}) and on the other hand the Casimir operator of $O(5)$ (\ref{casimir:O5}).  First we consider the internal commutation relations, in which case it is instructive to proceed by means of an example although the obtained result is generally valid.  Take e.g. the commutation relation $[T_{-\frac{1}{2}-\frac{1}{2}},T_{\frac{1}{2}\frac{1}{2}}]=\frac{1}{2}(X_0+Y_0)$, and sandwich it with the state $|vXM_XM_Y\rangle$
\begin{equation}
\eqalign{
\fl\langle vXM_XM_Y|T_{-\frac{1}{2}-\frac{1}{2}}T_{\frac{1}{2}\frac{1}{2}}|vXM_XM_Y\rangle-\langle vXM_XM_Y|T_{\frac{1}{2}\frac{1}{2}}T_{-\frac{1}{2}-\frac{1}{2}}|vXM_XM_Y\rangle\\=\case{1}{2}(M_X+M_Y).}
\end{equation}
At this point, we can insert a complete set of intermediate states $|v^\prime X^\prime M_X^\prime M_Y^\prime\rangle$ between the two generators.  
\begin{equation}
\eqalign{
\fl\sum_{v^\prime X^\prime M_X^\prime M_Y^\prime}\langle vXM_XM_Y|T_{-\frac{1}{2}-\frac{1}{2}}|v^\prime X^\prime M_X^\prime M_Y^\prime\rangle\langle v^\prime X^\prime M_X^\prime M_Y^\prime|T_{\frac{1}{2}\frac{1}{2}}|vXM_XM_Y\rangle\\
\fl-\sum_{v^\prime X^\prime M_X^\prime M_Y^\prime}\langle vXM_XM_Y|T_{\frac{1}{2}\frac{1}{2}}|v^\prime X^\prime M_X^\prime M_Y^\prime \rangle\langle v^\prime X^\prime M_X^\prime M_Y^\prime|T_{-\frac{1}{2}-\frac{1}{2}}|vXM_XM_Y\rangle\\
=\case{1}{2}(M_X+M_Y).}
\end{equation}
Due to symmetry considerations, a large amount of the matrix elements in the summation are identically zero.  First of all the $SU(2)_X\times SU(2)_Y$ bitensor character of $T$ dictates strict selection rules with respect to $X$, $M_X$ and $M_Y$.  As a result the summation over $X^\prime$, $M_X^\prime$ and $M_X^\prime$ is restricted to specific values which are completely governed by the Wigner-$3j$ symbol in (\ref{Tgenerators:awignereckart},\ref{Tgenerators:bwignereckart}).  Secondly, the components $T_{\mu\nu}$ of $T$ are $O(5)$ generators, which cannot alter the seniority quantum number $v$.  So, the summation over $v^\prime$ is reduced to one state $v^\prime=v$.\\
Once the restriction in the summation is carried out, it is convenient to apply the Wigner-Eckart theorem (\ref{Tgenerators:awignereckart},\ref{Tgenerators:bwignereckart}) and after some tedious algebra we obtain a relationship for the double reduced matrix elements
\begin{equation}\label{Tgenerators:relativereducedmatrixelement}
\eqalign{
\fl\frac{\langle vX||T||vX+\frac{1}{2}\rangle\langle vX+\frac{1}{2}||T||vX\rangle}{2X+2}-\frac{\langle vX||T||vX-\frac{1}{2}\rangle\langle vX-\frac{1}{2}||T||vX\rangle}{2X}
=\frac{(2X+1)^2}{2}.}
\end{equation}
The same procedure can be followed for the quadratic Casimir of $O(5)$.  Sandwiching equation~(\ref{casimir:O5}) with $|vXM_XM_Y\rangle$ yields
\begin{equation}
\eqalign{\fl \langle vXM_XM_Y|(T_{-\frac{1}{2}\frac{1}{2}}T_{\frac{1}{2}-\frac{1}{2}}+T_{\frac{1}{2}-\frac{1}{2}}T_{-\frac{1}{2}\frac{1}{2}}-T_{-\frac{1}{2}-\frac{1}{2}}T_{\frac{1}{2}\frac{1}{2}}-T_{\frac{1}{2}\frac{1}{2}}T_{-\frac{1}{2}-\frac{1}{2}})|vXM_XM_Y\rangle\\
=v(v+3)-4X(X+1).}
\end{equation}
By inserting again a complete set, applying the Wigner-Eckart theorem and making use of the previously derived relation (\ref{Tgenerators:relativereducedmatrixelement}), we obtain the result
\begin{equation}
\fl-4\langle vX||T||vX+\case{1}{2}\rangle\langle vX+\case{1}{2}||T||vX\rangle=(v-2X)(v+2X+3)(2X+1)(2X+2).
\end{equation}
This can slightly be rewritten, if one takes the Hermitian conjugate of the $T$ bitensor into consideration.
\begin{equation}
T_{\mu\nu}^\dag=(-1)^{\mu+\nu}T_{-\mu-\nu}.
\end{equation}
It can be proven that this leads towards the following expression for the double reduced matrix elements
\begin{equation}
\langle v,X||T||v,X+\case{1}{2}\rangle^\ast=-\langle v,X+\case{1}{2}||T||v,X\rangle.
\end{equation}
As a result, we can write 
\begin{eqnarray}
|\langle vX||T||vX+\case{1}{2}\rangle|^2&=\case{1}{4}(v-2X)(v+2X+3)(2X+1)(2X+2),\\
|\langle vX||T||vX-\case{1}{2}\rangle|^2&=\case{1}{4}(v-2X+1)(v+2X+2)(2X)(2X+1).
\end{eqnarray}
So the double reduced matrix elements are determined up to a phase.  Here we fix the relative sign of $\langle vX||T||vX+\frac{1}{2}\rangle$ and $\langle vX||T||vX-\frac{1}{2}\rangle$ to be opposite, as it is the only way to obtain eigenstates with real angular momentum $L$ in the physical basis (see section (\ref{section:rotation})).\\
Once that the double reduced matrix elements are determined, they can be plugged into equations (\ref{Tgenerators:awignereckart},\ref{Tgenerators:bwignereckart}), yielding the action of the bitensor $T$ components.
\begin{eqnarray}
\eqalign{\fl T_{\frac{1}{2}\frac{1}{2}}|vXM_XM_Y\rangle=\\
\fl\quad\frac{\sqrt{(X+M_X+1)(X+M_Y+1)(v-2X)(v+2X+3)}}{2\sqrt{(2X+1)(2X+2)}}|vX+\case{1}{2},M_X+\case{1}{2},M_Y+\case{1}{2}\rangle\label{Tgenerator:plusplus}\\
\fl\quad-\frac{\sqrt{(X-M_X)(X-M_Y)(v-2X+1)(v+2X+2)}}{2\sqrt{(2X)(2X+1)}}|vX-\case{1}{2},M_X+\case{1}{2},M_Y+\case{1}{2}\rangle,}\\
\eqalign{\fl T_{\frac{1}{2}-\frac{1}{2}}|vXM_XM_Y\rangle=\\
\fl\quad=\frac{\sqrt{(X+M_X+1)(X-M_Y+1)(v-2X)(v+2X+3)}}{2\sqrt{(2X+1)(2X+2)}}|vX+\case{1}{2},M_X+\case{1}{2},M_Y-\case{1}{2}\rangle\\
\fl\quad+\frac{\sqrt{(X-M_X)(X+M_Y)(v-2X+1)(v+2X+2)}}{2\sqrt{(2X)(2X+1)}}|vX-\case{1}{2},M_X+\case{1}{2},M_Y-\case{1}{2}\rangle,}\\
\eqalign{\fl T_{-\frac{1}{2}\frac{1}{2}}|vXM_XM_Y\rangle\\
\fl\quad=\frac{\sqrt{(X-M_X+1)(X+M_Y+1)(v-2X)(v+2X+3)}}{2\sqrt{(2X+1)(2X+2)}}|vX+\case{1}{2},M_X-\case{1}{2},M_Y+\case{1}{2}\rangle\\
\fl\quad+\frac{\sqrt{(X+M_X)(X-M_Y)(v-2X+1)(v+2X+2)}}{2\sqrt{(2X)(2X+1)}}|vX-\case{1}{2},M_X-\case{1}{2},M_Y+\case{1}{2}\rangle,}\\
\eqalign{\fl T_{-\frac{1}{2}-\frac{1}{2}}|vXM_XM_Y\rangle\\
\fl\quad=\frac{\sqrt{(X-M_X+1)(X-M_Y+1)(v-2X)(v+2X+3)}}{2\sqrt{(2X+1)(2X+2)}}|vX+\case{1}{2},M_X-\case{1}{2},M_Y-\case{1}{2}\rangle\\
\fl\quad-\frac{\sqrt{(X+M_X)(X+M_Y)(v-2X+1)(v+2X+2)}}{2\sqrt{(2X)(2X+1)}}|vX-\case{1}{2},M_X-\case{1}{2},M_Y-\case{1}{2}\rangle.}
\end{eqnarray}
From these expressions it is clearly seen that no representations can be constructed with $X>\frac{v}{2}$, as the representations must have a positive definite norm.  Combining these results with the standard quantum reduction rules for the $SU(2)$ group, we can label all basis states of a representation with fixed $v$ as follows
\begin{eqnarray}
\eqalign{X=0\dots v/2,\\
M_X=-X\dots X,\\
M_Y=-X\dots X.}
\end{eqnarray}
Figure~(\ref{figure:3Dmschema}) gives a visual interpretation of the reduction rules for the representation $v=2$.
\section{Matrix elements of collective variables}
%
%
The Hamiltonian describing a system undergoing quadrupole collective excitations contains a potential $V(\alpha)$, written in terms of the collective variables $\alpha_\mu$.  Indeed, it turns out that 
\begin{equation}
[\alpha\alpha]^{(2)}\cdot\alpha\sim\beta^3\cos3\gamma,
\end{equation}
can be considered as the building block of the $\gamma$ part of the potential $V(\beta,\gamma)$ in the intrinsic frame \cite{eisenberg:87}.  Therefore, matrix elements of $\alpha_\mu$ within a suitable basis are needed for the construction of the matrix representation of the Hamiltonian.  As the chosen framework in the present paper is the Cartan-Weyl natural basis, we proceed within this basis and show that all matrix elements can be calculated by means of an algebraic procedure, similar to the one proposed in the preceding section.\\
First, we need to establish the bitensor character of the collective variables $\alpha_\mu$ with respect to $SU(2)_X\times SU(2)_Y$.  Calculating the commutation relations of $\alpha_\mu$ with the $SU(2)$ generators (which is done most conveniently using the explicit expressions given in \ref{appendix:A}), we can summerize them as 
\begin{eqnarray}
&[X_{0},\alpha^{\lambda\lambda}_{\mu\nu}]=\mu \alpha^{\lambda\lambda}_{\mu\nu},\\
&[X_{\pm},\alpha^{\lambda\lambda}_{\mu\nu}]=\sqrt{(\lambda\mp\mu)(\lambda\pm\mu+1)}\alpha^{\lambda\lambda}_{\mu\pm1\nu},\\
&[Y_{0},\alpha^{\lambda\lambda}_{\mu\nu}]=\nu \alpha^{\lambda\lambda}_{\mu\nu},\\
&[Y_{\pm},\alpha^{\lambda\lambda}_{\mu\nu}]=\sqrt{(\lambda\mp\nu)(\lambda\pm\nu+1)}\alpha^{\lambda\lambda}_{\mu\nu\pm1},
\end{eqnarray}
where the 5 collective variables have been relabelled as follows
\begin{eqnarray}
\Big\{\alpha_2=\alpha_{\frac{1}{2}\frac{1}{2}}^{\frac{1}{2}\frac{1}{2}}, \alpha_1=\alpha_{-\frac{1}{2}\frac{1}{2}}^{\frac{1}{2}\frac{1}{2}}, \alpha_{-1}=\alpha_{\frac{1}{2}-\frac{1}{2}}^{\frac{1}{2}\frac{1}{2}}, \alpha_{-2}=\alpha_{-\frac{1}{2}-\frac{1}{2}}^{\frac{1}{2}\frac{1}{2}}\Big\},\\
\Big\{\alpha_0=\alpha_{00}^{00}\Big\}.
\end{eqnarray}
This clearly states that the 5 projections of $\alpha$ can be divided into the 4 components of a $\{\frac{1}{2}\frac{1}{2}\}$ bispinor and a single biscalar, according to Racah \cite{racah:42}.  We can again define double reduced matrix elements
\begin{equation}\label{variables:wignereckart}
\eqalign{
\langle v X M_XM_Y|\alpha^{\lambda\lambda}_{\mu\nu}|v^\prime X^\prime M_X^\prime M_Y^\prime\rangle\\
=(-)^{k}\wigner{X}{\lambda}{X^\prime}{-M_X}{\mu}{M_X^\prime}\wigner{X}{\lambda}{X^\prime}{-M_Y}{\nu}{M_Y^\prime}\langle vX||\alpha^\lambda||v^\prime X^\prime\rangle,}
\end{equation}
with $k=2X-M_X-M_Y$.  It is noteworthy that, contrary to the matrix elements of the generators $T_{\mu\nu}$, $v^\prime$ is not necessarily equal to $v$.  To obtain explicit expressions for the double reduced matrix elements, we start from the commutation relations
\begin{eqnarray}
[T_{\mu\nu},\alpha_{\mu^\prime\nu^\prime}^{\frac{1}{2}\frac{1}{2}}]=\frac{(-)^{(\mu+\nu)}}{\sqrt{2}}\delta_{-\mu\mu^\prime}\delta_{-\nu\nu^\prime}\alpha_{00}^{00}\label{matelem:commute1},\\{}
[T_{\mu\nu},\alpha_{00}^{00}]=\frac{1}{\sqrt{2}}\alpha_{\mu\nu}^{\frac{1}{2}\frac{1}{2}}\label{matelem:commute2},\\{}
[\alpha_{\mu\nu}^{\lambda\lambda},\alpha_{\mu^\prime\nu^\prime}^{\lambda^\prime\lambda^\prime}]=0.\label{matelem:commute3}
\end{eqnarray}
First a relationship between $\langle vX||\alpha^0||v^\prime X^\prime\rangle$ and $\langle vX||\alpha^\frac{1}{2}||v^\prime X^\prime\rangle$ needs to be established.  This can be accomplished using the commutation relations (\ref{matelem:commute1}).   We consider the specific case $[T_{\frac{1}{2}\frac{1}{2}},\alpha_{\frac{1}{2}\frac{1}{2}}^{\frac{1}{2}\frac{1}{2}}]=0$ and construct the following matrix elements
\begin{eqnarray}
\langle vX\pm\case{1}{2},M_X+\case{1}{2},M_Y+\case{1}{2}|[T_{\frac{1}{2}\frac{1}{2}},\alpha_{\frac{1}{2}\frac{1}{2}}^{\frac{1}{2}\frac{1}{2}}]|v^\prime X\mp \case{1}{2},M_X-\case{1}{2},M_Y-\case{1}{2}\rangle,\label{matelem:Talphasandwich1}\\
\langle vXM_XM_Y|[T_{\frac{1}{2}\frac{1}{2}},\alpha_{\frac{1}{2}\frac{1}{2}}^{\frac{1}{2}\frac{1}{2}}]|v^\prime XM_X-1,M_Y-1\rangle,\label{matelem:Talphasandwich2}
\end{eqnarray}
in terms of the double reduced matrix elements.  This can be achieved by inserting a complete set of basis states between the generator $T_{\frac{1}{2}\frac{1}{2}}$ and the variable $\alpha_{\frac{1}{2}\frac{1}{2}}^{\frac{1}{2}\frac{1}{2}}$, then making use of the $T$ matrix elements (\ref{Tgenerator:plusplus}) obtained in the previous section and the Wigner-Eckart theorem (\ref{variables:wignereckart}).  The outcome of these tedious although straightforward calculations are, respectively for (\ref{matelem:Talphasandwich1}) (with minus and positive sign) and (\ref{matelem:Talphasandwich2})
\begin{eqnarray}
\eqalign{\langle vX-\case{1}{2}||\alpha^\frac{1}{2}||v^\prime X\rangle\frac{\sqrt{(v^\prime-2X)(v^\prime+2X+3)}}{\sqrt{2X}}\\
\quad-\langle vX||\alpha^\frac{1}{2}||v^\prime X+\case{1}{2}\rangle\frac{\sqrt{(v-2X+1)(v+2X+2)}}{\sqrt{2X+2}}=0,}\label{matelem:rule1}\\
\eqalign{\langle vX+\case{1}{2}||\alpha^\frac{1}{2}||v^\prime X\rangle\frac{\sqrt{(v^\prime-2X+1)(v^\prime+2X+2)}}{\sqrt{2X+2}}\\
\quad-\langle vX||\alpha^\frac{1}{2}||v^\prime X-\case{1}{2}\rangle\frac{\sqrt{(v-2X)(v+2X+3)}}{\sqrt{2X}}=0,}\label{matelem:rule2}\\
\eqalign{\frac{\langle vX+\frac{1}{2}||\alpha^{\frac{1}{2}}||v^\prime X\rangle}{\sqrt{(v-2X)(v+2X+3)}}\left[\frac{(v^\prime+1)(v^\prime+2)}{2X}-\frac{(v+1)(v+2)}{2X+2}\right]\\
\quad+\frac{\langle vX||\alpha^{\frac{1}{2}}||v^\prime X+\frac{1}{2}\rangle}{\sqrt{(v^\prime-2X)(v^\prime+2X+3)}}\left[\frac{(v+1)(v+2)}{2X}-\frac{(v^\prime+1)(v^\prime+2)}{2X+2}\right]=0.\label{matelem:rule3}}
\end{eqnarray}
The same procedure can be repeated for the commutation relation $[T_{\frac{1}{2}\frac{1}{2}},\alpha_{-\frac{1}{2}\frac{1}{2}}^{\frac{1}{2}\frac{1}{2}}]=0$.  We obtain again (\ref{matelem:rule1})~and~(\ref{matelem:rule2}), accompanied by the following expression
\begin{equation}
\eqalign{\fl\frac{\langle vX+\frac{1}{2}||\alpha^{\frac{1}{2}}||v^\prime X\rangle}{\sqrt{(v-2X)(v+2X+3)}}\left[(v-2X)(v+2X+3)-(v^\prime -2X+1)(v^\prime+2X+2)\right]\\
\fl\quad+\frac{\langle vX||\alpha^{\frac{1}{2}}||v^\prime X+\frac{1}{2}\rangle}{\sqrt{(v^\prime-2X)(v^\prime+2X+3)}}\left[(v-2X+1)(v+2X+2)-(v^\prime-2X)(v^\prime +2X+3)\right]=0}.\label{matelem:rule4}
\end{equation}
Combining (\ref{matelem:rule3}) with (\ref{matelem:rule4}) gives a homogeneous set of two equations in two variables $\langle vX+\frac{1}{2}||\alpha^{\frac{1}{2}}||v^\prime X\rangle$ and $\langle vX||\alpha^{\frac{1}{2}}||v^\prime X+\frac{1}{2}\rangle$, rendering the trivial zero solution, unless the determinant of the associated matrix identically vanishes.  This is only possible when $v^\prime=v\pm1$, which proves the common knowledge that $\alpha$ forms an $O(5)$-tensor of rank 1.\\
Finally, we repeat the procedure for $[T_{\frac{1}{2}\frac{1}{2}},\alpha_{-\frac{1}{2}-\frac{1}{2}}^{\frac{1}{2}\frac{1}{2}}]=-\frac{1}{\sqrt{2}}\alpha_{00}^{00}$.  Besides (\ref{matelem:rule1}) and (\ref{matelem:rule2}), we obtain the expression
\begin{equation}
\eqalign{\fl\frac{\langle vX+\frac{1}{2}||\alpha^{\frac{1}{2}}||v^\prime X\rangle}{\sqrt{(v-2X)(v+2X+3)}}\left[X(v^\prime+1)(v^\prime+2)-(X+1)(v+1)(v+2)+2(2X+1)^2\right]\\
\fl\quad+\frac{\langle vX||\alpha^{\frac{1}{2}}||v^\prime X+\frac{1}{2}\rangle}{\sqrt{(v^\prime-2X)(v^\prime+2X+3)}}\left[X(v+1)(v+2)-(X+1)(v^\prime+1)(v^\prime+2)+2(2X+1)^2\right]\\
\quad=-2\sqrt{2}\sqrt{(2X+1)(2X+2)}\langle vX||\alpha^0||v^\prime X\rangle.}\label{matelem:rule5}
\end{equation}
Solving the set of equations (\ref{matelem:rule4}) and (\ref{matelem:rule5}) (or equivalently (\ref{matelem:rule3}) and (\ref{matelem:rule5})) results in expressions of all possible double reduced matrix elements of $\alpha^{\frac{1}{2}}$ as a function of the double reduced matrix elements of $\alpha^{0}$.\\
From this point on, we will explicitly take into account that the $\alpha$ variable connects representations with $\Delta v=1$, omitting the other matrix elements which are identically zero.  For $v^\prime=v+1$ we obtain
\begin{eqnarray}
\fl\langle v,X+\case{1}{2}||\alpha^{\frac{1}{2}}||v+1,X\rangle&=-\frac{1}{\sqrt{2}}\sqrt{\frac{2X+2}{2X+1}}\sqrt{\frac{v-2X}{v+2X+3}}\langle vX||\alpha^0||v+1,X\rangle\label{matelem:rule6a},\\
\fl\langle v,X||\alpha^{\frac{1}{2}}||v+1,X+\case{1}{2}\rangle&=\frac{1}{\sqrt{2}}\sqrt{\frac{2X+2}{2X+1}}\sqrt{\frac{v+2X+4}{v-2X+1}}\langle vX||\alpha^0||v+1,X\rangle\label{matelem:rule6b},
\end{eqnarray}
and for $v^\prime=v-1$
\begin{eqnarray}
\fl\langle v,X+\case{1}{2}||\alpha^{\frac{1}{2}}||v-1,X\rangle&=\frac{1}{\sqrt{2}}\sqrt{\frac{2X+2}{2X+1}}\sqrt{\frac{v+2X+3}{v-2X}}\langle vX||\alpha^0||v-1,X\rangle\label{matelem:rule7a},\\
\fl\langle v,X||\alpha^{\frac{1}{2}}||v-1,X+\case{1}{2}\rangle&=-\frac{1}{\sqrt{2}}\sqrt{\frac{2X+2}{2X+1}}\sqrt{\frac{v-2X-1}{v+2X+2}}\langle vX||\alpha^0||v-1,X\rangle.\label{matelem:rule7b}
\end{eqnarray}
So, we only have to determine the biscalar double reduced matrix elements.  Although the commutation relations~(\ref{matelem:commute3}) seem trivial, they are convenient in the derivation of the $\{00\}$ double reduced matrix elements.   If we consider only the non-trivial commutation relations for which $\lambda=\frac{1}{2}$ and $\lambda^\prime=\frac{1}{2}$, and apply again the same procedure which has been used throughout the present paper, we obtain the following result
\begin{equation}
\eqalign{\frac{\sum_{v^\prime}\langle vX||\alpha^{\frac{1}{2}}||v^\prime X+\frac{1}{2}\rangle\langle v^\prime X+\frac{1}{2}||\alpha^{\frac{1}{2}}||vX\rangle}{2X+2}\\
\qquad-\frac{\sum_{v^\prime}\langle vX||\alpha^{\frac{1}{2}}||v^\prime X-\frac{1}{2}\rangle\langle v^\prime X-\frac{1}{2}||\alpha^{\frac{1}{2}}||vX\rangle}{2X}=0.}\label{matelem:rule8}
\end{equation}
Now, taking all derived expressions (\ref{matelem:rule1}),(\ref{matelem:rule2}) and (\ref{matelem:rule6a}) to (\ref{matelem:rule7b}) into account, we can rewrite the relation~(\ref{matelem:rule8}) as
\begin{equation}
\eqalign{\frac{(2v+5)}{(v-2X+1)(v+2X+3)}\langle vX||\alpha^{0}||v+1 X\rangle\langle v+1X||\alpha^{0}||vX\rangle\\
\qquad=\frac{(2v+1)}{(v+2X+2)(v-2X)}\langle vX||\alpha^{0}||v-1 X\rangle\langle v-1X||\alpha^{0}||vX\rangle.}\label{matelem:rule9}
\end{equation}
This relation differs from the previously derived expressions with respect to the quantum numbers.  The expressions (\ref{matelem:rule1}) to (\ref{matelem:rule7b}) relate matrix elements with different $X$ connections, though the seniority connection ($v$ to $v^\prime=v\pm1$) was fixed.  Now, (\ref{matelem:rule9}) relates matrix elements with different seniority connection, leaving the $X$ quantum number unaltered.\\
At last, in order to obtain explicit expressions, we return to the geometry of the problem.  It has been mentioned earlier that the operator $\alpha\cdot\alpha$ commutes with all the generators of $O(5)$, making it an $O(5)$ scalar. Therefore, this operator can be treated as a constant with respect to the $O(5)$ scheme.  We call this constant $\beta^2$, referring to the radial deformation parameter in (\ref{variables:intrinsicframe}).  As a consequence, we can write  
\begin{equation}\label{matelem:betasquared}
\langle vXM_XM_Y|\alpha\cdot\alpha|vXM_XM_Y\rangle=\beta^2.\label{matelem:rule10}
\end{equation}
The procedure used culminates into closed expressions of the matrix elements.  By inserting a complete set of basis states between the variables of (\ref{matelem:rule10}), we can rewrite this expression in terms of double reduced matrix elements.  
\begin{equation}
\eqalign{\beta^2=\frac{1}{(2X+1)^2}\sum_{v^\prime}&[\langle vX||\alpha^0||v^\prime X\rangle\langle v^\prime X||\alpha^0||v X\rangle\\
&+\langle vX||\alpha^{\frac{1}{2}}||v^\prime X+\case{1}{2}\rangle\langle v^\prime X+\case{1}{2}||\alpha^{\frac{1}{2}}||v X\rangle\\
&+\langle vX||\alpha^{\frac{1}{2}}||v^\prime X-\case{1}{2}\rangle\langle v^\prime X-\case{1}{2}||\alpha^{\frac{1}{2}}||v X\rangle].}
\end{equation}
All these different matrix elements can be reduced to a single one by means of the reduction rules (\ref{matelem:rule1}) to (\ref{matelem:rule9}).  As a result, we obtain
\begin{eqnarray}
\fl\langle v,X||\alpha^0||v+1,X\rangle\langle v+1,X||\alpha^0||v,X\rangle&=\frac{(v-2X+1)(v+2X+3)}{(2v+3)(2v+5)}(2X+1)^2\beta^2,\\
\fl\langle v,X||\alpha^0||v-1,X\rangle\langle v-1,X||\alpha^0||v,X\rangle&=\frac{(v-2X)(v+2X+2)}{(2v+1)(2v+3)}(2X+1)^2\beta^2.
\end{eqnarray}
Taking into account that $\alpha^\dag_0=\alpha_0$, we can write
\begin{equation}
\langle vX||\alpha^0||v^\prime X\rangle^\ast=\langle v^\prime X||\alpha^0||vX\rangle,
\end{equation}
and summerize
\begin{eqnarray}
|\langle v,X||\alpha^0||v+1,X\rangle|^2&=\frac{(v-2X+1)(v+2X+3)}{(2v+3)(2v+5)}(2X+1)^2\beta^2,\\
|\langle v,X||\alpha^0||v-1,X\rangle|^2&=\frac{(v-2X)(v+2X+2)}{(2v+1)(2v+3)}(2X+1)^2\beta^2,
\end{eqnarray}
which is equivalent to
\begin{eqnarray}
\langle v,X||\alpha^0||v+1,X\rangle=\sqrt{\frac{(v-2X+1)(v+2X+3)}{(2v+3)(2v+5)}}(2X+1)\beta,\\
\langle v,X||\alpha^0||v-1,X\rangle=\sqrt{\frac{(v-2X)(v+2X+2)}{(2v+1)(2v+3)}}(2X+1)\beta,
\end{eqnarray}
as $\alpha_0$ is a hermitian operator.\\
Making use of the reduction rules (\ref{matelem:rule6a}-\ref{matelem:rule7b}), we can construct all double reduced matrix elements of the $\alpha$ variable.  Taking the appropriate Wigner-$3j$ coefficients into account, the total matrix elements of the $\alpha$ variable can easily be derived.  As an example we evaluate the matrix element
\begin{equation}
\eqalign{\langle v&XM_XM_Y|\alpha^{00}_{00}|v^\prime XM_XM_Y\rangle\\
&=(-)^{k}\wigner{X}{0}{X}{-M_X}{0}{M_X}\wigner{X}{0}{X}{-M_Y}{0}{M_Y}\langle vX||\alpha^0||v^\prime,X\rangle,}
\end{equation}
(with $k=2X-M_X-M_Y$).  This leads to the closed expression
\begin{eqnarray}
\langle vXM_XM_Y|\alpha^{00}_{00}|v+1,XM_XM_Y\rangle&=\beta\sqrt{\frac{(v-2X+1)(v+2X+3)}{(2v+3)(2v+5)}},\\
\langle vXM_XM_Y|\alpha^{00}_{00}|v-1,XM_XM_Y\rangle&=\beta\sqrt{\frac{(v-2X)(v+2X+2)}{(2v+1)(2v+3)}}.
\end{eqnarray}
There is a subtlety involved with equation (\ref{matelem:betasquared}).  $\beta^2$ can be either be regarded as the radial variable in the 5-dimensional Euclidean space, which is a constant by definition under rotations of the $O(5)$ orthogonal group, or it can be recognized as a generator of the aforementioned $SU(1,1)$ algebra.  In the latter scheme, the $O(5)$ Hilbert space needs to be extended to incorporate this $SU(1,1)$ basis.  Then, it is more convenient to move over to a boson creation and annihilation realization
\begin{equation}
b_\mu^\dag=\case{1}{\sqrt{2}}(\sqrt{k}\alpha_\mu+\case{i}{\sqrt{k}\hbar}\pster_\mu),\qquad \tilde{b}_\mu=\case{1}{\sqrt{2}}(\sqrt{k}\alpha_\mu-\case{i}{\sqrt{k}\hbar}\pster_\mu),
\end{equation}
with $[b_\mu,b_\nu^\dag]=\delta_{\mu\nu}$ and $\tilde{b}_\mu=(-1)^\mu b_{-\mu}$, as it gives immediately rise to the $SU(1,1)$ algebra spanned by
\begin{equation}
B_+=\case{1}{2}b^\dag\cdot b^\dag,\quad B_-=\case{1}{2}\tilde{b}\cdot \tilde{b},\quad B_0=\case{1}{4}(b^\dag\cdot\tilde{b}+\tilde{b}\cdot b^\dag),
\end{equation}
which is closely connected to the Hamiltonian of a spherical harmonic oscillator via the Cartan operator $B_0$.  A recent study \cite{rowe:05a} has proven that an extension to a Davidson based $SU(1,1)$ algebra has a major advantage over the spherical basis in the sense that the $\beta$ deformation of atomic nuclei is naturally included in the basis representations, leading to a higher convergency speed when dealing with actual deformations in collective nuclear model calculations.  Therefore, we prefer to regard $\beta^2$ as a constant over $O(5)$ rotations, as no bosonic representations of the Davidson $SU(1,1)$ are known.  Moreover recent developments \cite{rowe:05b} in the framework of the factorization method \cite{infeld:51} have shown that it is possible to extract the needed $\beta$ matrix elements by means of generalized raising and lowering operators for a Davidson type of potentials.\\
\section{Rotation to the physical basis}\label{section:rotation}
\begin{figure}[!htb]
\begin{center}
\includegraphics{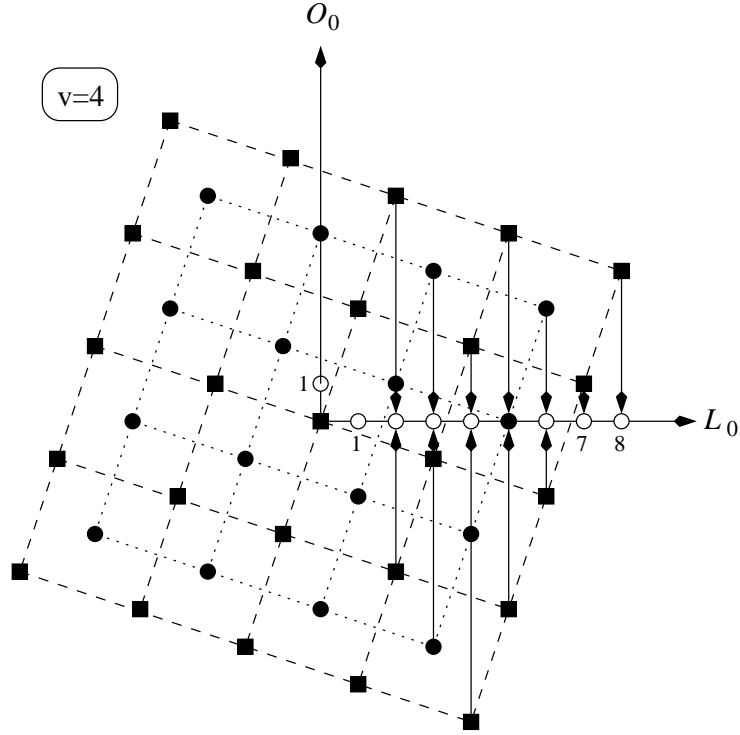}
\end{center}
\caption{Projections of the $v=4$ irreps on the $L_0$ axis, connecting all basis states for which $M=M_X+3M_Y$ is equal.  The line most on the right only projects the $M_X=v/2, M_Y=v/2$ state on the $L_0$ axis, resulting in an $M=2v$ state.  For $v=4$ the maximal $L_0$ projection corresponds to $M=8$}\label{figure:MschemaX}
\end{figure}

\begin{figure}[!htb]
\begin{center}
\includegraphics{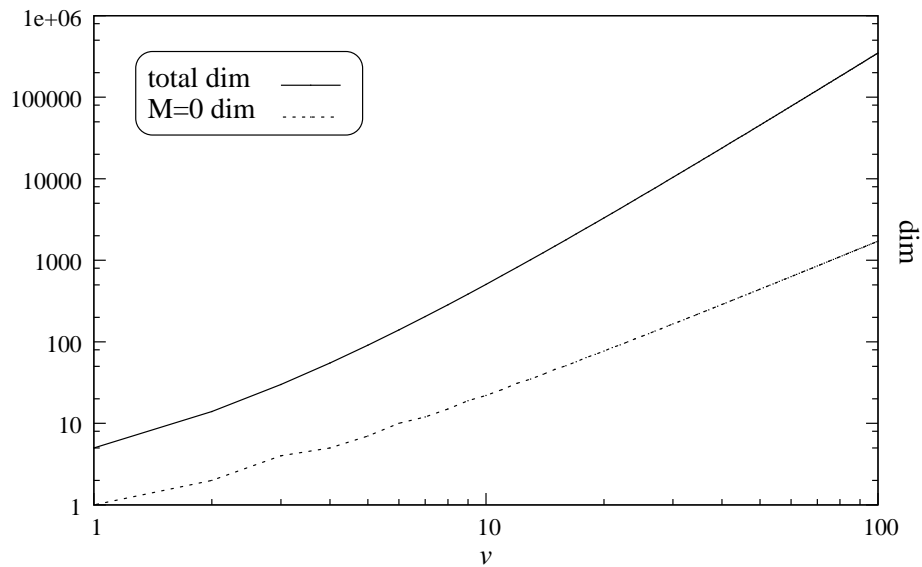}
\end{center}
\caption{Total dimension of a representations with a given seniority $v$, compared to the dimension of the subspace of states with $M=0$.  It is clear that the dimension of the $M=0$ subspace increases quadratically with increasing seniority, however it remains feasible for realistic calculations.   For $v=100$, the dimension of $M=0$ space is $1717$, while the total dimension is $348 551$.}\label{figure:dimv}
\end{figure}

%
%
The drawback of the Cartan-Weyl reduction is that its basis is not naturally compatible with the physical angular momentum quantum number $L$ which emerges from experimental energy spectra.  This comes from the fact that $L\cdot L$ does not commute with $X^2$, implying that a basis diagonalizing both operators is non-existent.  Therefore, a rotation from the natural group chain $O(5)\supset O(4)\cong SU(2)\times SU(2)$ to the physical chain $O(5)\supset O(3)\supset O(2)$ is needed.  Fortunately, the Casimir operator $L_0$ associated with the physical $O(2)$ group is diagonal in the Cartan-Weyl basis, leaving $L\cdot L$ the only operator to diagonalize.
\begin{equation}
L\cdot L=L_0^2+\case{1}{2}(L_{+}L_{-}+L_{-}L_{+}).
\end{equation}
Rewriting the $O(3)$ generators in terms of the Cartan-Weyl generators gives
\begin{eqnarray}
L_\pm&=2X_\pm+\sqrt{12}T_{\mp\frac{1}{2}\pm\frac{1}{2}},\\
L_0&=X_0+3Y_0,
\end{eqnarray}
so that $L\cdot L$ can be written as
\begin{equation}
\eqalign{L\cdot L&=4X^2 -3[(X_0-3Y_0+\case{1}{2})(X_0+Y_0+\case{1}{2})-\case{1}{4}]\\
&\quad +4\sqrt{3}[T_{-\frac{1}{2}\frac{1}{2}}X_{-}+T_{\frac{1}{2}-\frac{1}{2}}X_{+}]+12T_{\frac{1}{2}-\frac{1}{2}}T_{-\frac{1}{2}\frac{1}{2}}.}
 \end{equation}
The action of all generators are known in the natural basis (see section~\ref{section:representations}), so the matrix elements of a matrix representation can easily be calculated.  The dimension of the matrix is governed by the seniority quantum number since the generators involved in the expression $L\cdot L$ cannot alter the $O(5)$ quantum number, which means that there is an associated matrix with every $v$.  This matrix is even further reducible if one takes the $L_0$ operator into account.  Since $L_0$ can be written as $L_0=X_0+3Y_0$, it is immediately diagonal in the Cartan-Weyl basis, making $M=M_X+3M_Y$ a good quantum number. As a consequence, the total matrix representation of $L\cdot L$ can be divided in separate sub-matrices with distinct $M$ quantum number.  The possible basis states $|vXM_XM_Y\rangle$ spanning the sub-matrices with $M=M_X+3M_Y$ can easily be recognized in the tilted weight diagrams (Figure~\ref{figure:MschemaX}).  The diagram is tilted with respect to the angular momentum operator $L_0$, so that the vertical projection of every basis state immediately gives the $L_0$ component.  As a result, all basis states, lying on the same 
vertical projection line form a subspace of states for which $M=M_X+3M_Y$ holds.\\
Although the rotation from the natural towards the physical basis corresponds to a standard diagonalization problem, it is useful to study some specific cases.  It is readily seen from figure~\ref{figure:MschemaX} that the projection $M=2v$ can only be constructed from one single basis state $|v,X=\frac{v}{2},M_X=\frac{v}{2},M_Y=\frac{v}{2}\rangle$, as there is only one projection state.  Therefore, the matrix representation is one dimensional
\begin{equation}
\langle v,\case{v}{2}(\case{v}{2}\case{v}{2})|L\cdot L|v,\case{v}{2}(\case{v}{2}\case{v}{2})\rangle=2v(2v+1).
\end{equation}
The same is valid for $M=2v-1$.  The only basis state with this $M$ projection is $|v,X=\frac{v}{2},M_X=\frac{v}{2}-1,M_Y=\frac{v}{2}\rangle$, giving the same eigenvalue $2v(2v+1)$.  For $M=2v-2$, there are two different states: $|v,X=\frac{v}{2},M_X=\frac{v}{2}-2,M_Y=\frac{v}{2}\rangle$ and $|v,X=\frac{v}{2}-\frac{1}{2},M_X=\frac{v}{2}-\frac{1}{2},M_Y=\frac{v}{2}-\frac{1}{2}\rangle$.  
which gives a 2 dimensional matrix with eigenvalues and accompanying eigenvectors
\begin{eqnarray}
\lambda_+=2v(2v+1) & \rightarrow \left(\sqrt{\case{4(v-1)}{4v-1}},\sqrt{\case{3}{4v-1}}\right),\\
\lambda_-=(2v-2)(2v-1) & \rightarrow \left(\sqrt{\case{3}{4v-1}},-\sqrt{\case{4(v-1)}{4v-1}}\right).
\end{eqnarray}
So it is clear that the associated eigenvector of $\lambda_+$ belongs to the $L=2v$ multiplet while the eigenvector of $\lambda_-$ will be the heighest $M$ state of the $L=2v-2$ multiplet.  Basically, this procedure can be repeated up to $M=0$ by means of a symbolic mathematical computer program or by means of numerical procedures.\\
Finally we discuss the dimension of the matrix representations, as they are important in actual calculations.  The total number of basis states within a representation $v$ can be determined as
\begin{equation}
\sum_{X=0}^{v/2}(2X+1)^2=\frac{1}{6}(2v+3)(v+2)(v+1).
\end{equation}
However, the dimensions of the $M=0$ sub-blocks are lower than the total $v$ representation space.  These dimensions can be calculated according to the following formula, depending whether $v$ is even or odd.  If we define $X_m=v/2$, the number of $M=0$ projections is then given by
\begin{equation}
\eqalign{3(X_m|3)^2&+[X_m\textrm{mod}3+1][2(X_m|3)+1]+3[(X_m+1)|3-1][(X_m+1)|3]\\
&+2[(X_m+1)|3][(X_m+1)\textrm{mod}3+1],}
\end{equation}
for even $v$. For odd $v$ we obtain
\begin{equation}
\eqalign{3[(X_m&+\case{3}{2})|3-1][(X_m+\case{3}{2})|3]+2[(X_m+\case{3}{2})|3][(X_m+\case{3}{2})\textrm{mod}3+1]\\
&+3[(X_m-\case{1}{2})|3]^2+[(X_m-\case{1}{2})\textrm{mod}3+1][2((X_m-\case{1}{2})|3)+1].}
\end{equation}
These two dimension formulas are plotted in figure~\ref{figure:dimv}.  
From this figure, it is clear that the dimension of the $M=0$ subspace stays reasonable with respect to modern computation standards, as long as relatively low-order seniorities are considered.  Anyhow, when performing realistic calculations, the transformation from natural to physical basis does not need to be repeated for every calculation, as the rotation is independent of the specific physical system (Hamiltonian) under study.  In practical calculations, the rotation only has to be carried out once and stored for later use.
\section{Conclusions and outlook}\label{section:conclusions}
%
%
Collective modes of motions have proven to be very important in atomic nuclei, away from the shell-closures.  Therefore it is of major interest to construct schematic Hamiltonians in the significant degrees of freedom.  Throughout the last decades much attention has been given to solving the Schr\"odinger equation for general potential energy surfaces in the collective variables.  This resulted in a number of techniques, based on combinations of analytical and algebraic considerations.  The present manuscript adds a method which is completely algebraic in the sense that no normalized highest weight states need to be constructed.  As a matter of fact, although the $\gamma$-rotational structure of the collective model is completely contained in the $O(5)$ subgroup of $SU(1,1)\times O(5)$, no explicit representations in terms of $\gamma$ had to be constructed to obtain matrix elements of the collective variables, relevant for general potential energy surfaces.  This suggests that the proposed technique can be extended to higher rank algebras, such as the $O(7)$ orthogonal group, emerging from octupole degrees of freedom in atomic nuclei.\\
Now the theoretical framework is set, it is interesting to study to what extend the geometrical model can be applied to the collective behaviour of atomic nuclei with respect to the recent developments in exotic nuclei.  This will be the subject of further investigations.
\section*{Acknowledgments}
%
%
The authors wish to thank Piet Van Isacker and John Wood for interesting discussions and suggestions.  Financial support from the University of Ghent and the ''FWO-Vlaanderen'' that made this research possible is acknowledged.  Also the Interuniversity Attraction Pool (IUAP) under project P5/07 is acknowledged for financial support.
\appendix
%
%
\section{Explicit expressions of the $O(5)$ generators in the Cartan-Weyl basis}\label{appendix:A}
%
%
The generators $L_M$ and $O_M$ can explicitly be expressed in terms of the collective variables and their canonic conjugate momenta according to the definition (\ref{LMOM:definition})
\begin{equation}
\eqalign{
L_M&=-\frac{i\sqrt{10}}{\hbar}[\alpha\pi^\ast]^{1}_M
=-\frac{i\sqrt{10}}{\hbar}\sum_{\mu\mu^\prime}\langle 2\mu2\mu^\prime|1M\rangle\alpha_\mu\pi^\ast_{\mu^\prime},\\
O_M&=-\frac{i\sqrt{10}}{\hbar}[\alpha\pi^\ast]^{3}_M
=-\frac{i\sqrt{10}}{\hbar}\sum_{\mu\mu^\prime}\langle 2\mu2\mu^\prime|3M\rangle\alpha_\mu\pi^\ast_{\mu^\prime},}
\end{equation}
where $\langle j_1m_1j_2m_2|j_3m_3\rangle$ are the commonly known Clebsch Gordan coefficients. 
Taking the rotation to the Cartan representation into account (\ref{generators:rotationtocartan}), explicit and relatively simple expressions for the generators can be obtained
\begin{equation}\label{appendix:explicit}
\eqalign{
X_{+}=\frac{i}{\hbar}(\alpha_2\pster_{-1}-\alpha_{-1}\pster_{2}), & Y_{+}=\frac{i}{\hbar}(\alpha_2\pster_{1}-\alpha_{1}\pster_{2}),\\
X_{-}=\frac{-i}{\hbar}(\alpha_1\pster_{-2}-\alpha_{-2}\pster_{1}), & Y_{-}=\frac{-i}{\hbar}(\alpha_{-1}\pster_{-2}-\alpha_{-2}\pster_{-1}),\\
X_0=\frac{-i}{2\hbar}(\alpha_2\pster_{-2}+\alpha_1\pster_{-1}-\alpha_{-1}\pster_{1}-\alpha_{-2}\pster_{2}), &{}\\ Y_0=\frac{-i}{2\hbar}(\alpha_2\pster_{-2}-\alpha_1\pster_{-1}+\alpha_{-1}\pster_{1}-\alpha_{-2}\pster_{2}), & {}\\
T_{\frac{1}{2}\frac{1}{2}}=\frac{-i}{\hbar\sqrt{2}}(\alpha_2\pster_{0}-\alpha_{0}\pster_{2}), & 
T_{-\frac{1}{2}\frac{1}{2}}=\frac{-i}{\hbar\sqrt{2}}(\alpha_1\pster_{0}-\alpha_{0}\pster_{1}),\\
T_{\frac{1}{2}-\frac{1}{2}}=\frac{-i}{\hbar\sqrt{2}}(\alpha_{-1}\pster_{0}-\alpha_{0}\pster_{-1}), &
T_{-\frac{1}{2}-\frac{1}{2}}=\frac{-i}{\hbar\sqrt{2}}(\alpha_{-2}\pster_{0}-\alpha_{0}\pster_{-2}).}
\end{equation}
\section*{References}
%
%
\bibliography{debaerdemacker_quadrupolevariables}
\bibliographystyle{iopart-num}
\end{document}